\documentclass[10pt, aps, prl, superscriptaddress, noshowkeys, 
               twocolumn, floatfix, preprintnumbers]{revtex4-1}
\usepackage{amsmath, amssymb, amsfonts}
\usepackage{graphicx}
\usepackage{color}
\usepackage{times}
\newcommand{\bk}{{\bf{k}}}
\newcommand{\bp}{{\bf{p}}}
\newcommand{\bq}{{\bf{q}}}
\newcommand{\by}{{\bf{y}}}
\newcommand{\ud}{\text{d}}

\begin{document}
\title{Excitons in Graphene and the Influence of the Dielectric Environment}

\author{J. H. Gr{\"o}nqvist}
\affiliation{Department of Physics,
\AA bo Akademi University, 20500 Turku, Finland}
\affiliation{Department of Physics and Material Sciences Center, 
Philipps University Marburg, Renthof 5, D-35032 Marburg, Germany}

\author{T. Stroucken}
\affiliation{Department of Physics and Material Sciences Center, 
Philipps University Marburg, Renthof 5, D-35032 Marburg, Germany}

\author{G. Bergh{\"a}user}
\affiliation{Department of Physics and Material Sciences Center, 
Philipps University Marburg, Renthof 5, D-35032 Marburg, Germany}

\author{S.W. Koch}
\affiliation{Department of Physics and Material Sciences Center, 
Philipps University Marburg, Renthof 5, D-35032 Marburg, Germany}

\begin{abstract}
The exciton Wannier equation for graphene is solved for
different background dielectric constants. It is shown that freestanding
graphene features strong Coulomb effects with a very large exciton
binding energy exceeding $3\,$eV. A second-order transition
to a weak Coulomb regime is found if the effective background dielectric constant
exceeds a critical value. All bound-state solutions vanish for epitaxial graphene 
on a substrate with large background permittivity, such as SiC.
\end{abstract}
\date{\today}
\maketitle

Graphene is an effectively two-dimensional (2D) system with unusual electronic and optical 
characteristics.
It is widely believed that the electronic properties result from the $sp^2$ hybridization 
between one $s$ and two $p$ orbitals of the constituting carbon atoms, leading to the typical 
planar honeycomb lattice.
The remaining $p$ orbital forms a half-filled $\pi$ band responsible for the electronic and 
optical properties \cite{Wallace47, Semenoff84, Castro-Neto2009}.
Using the tight binding(TB) approximation with nearest neighbor hopping only, it was first shown 
by Wallace \cite{Wallace47} that the hexagonal symmetry of the honeycomb lattice leads 
to the formation of symmetric electron and hole bands that touch each other at two nodes. 
In the vicinity of these so-called Dirac points, the dispersion is linear and the density 
of states vanishes.
According to this picture, which is supported by many experiments 
\cite{Novoselov2005, Zhang2005, Deacon2007, Jiang2007, Zhou2006, Bostwick2007}, 
graphene behaves as a semimetal whose massless Dirac electrons have a Fermi velocity
$v_F\approx 10^6 \text{m/s}$.

Whereas the existence of quasi-relativistic electrons is generally accepted, much less agreement 
exists concerning the role and the consequences of Coulomb interaction effects in graphene 
\cite{Gonzalez99, Sheehy2007, Fritz2008, Sinner2010, Khveshchenko2001, Khveshchenko2006, 
Juricic2009, Drut2009, Drut2009b, Yang2009, Malic2010, 
Guclu2010, Reed2010, Min2008, Berman2008a, Berman2008b}.
Studies based on a renormalization-group analysis predict
a stable semimetallic ground state \cite{Gonzalez99,Sheehy2007,Fritz2008,Sinner2010},
however, nonperturbative methods yield a semimetal to excitonic insulator transition at 
sufficiently high coupling strengths
\cite{Khveshchenko2001, Khveshchenko2006, Juricic2009, Drut2009, Drut2009b}.
 
As a convenient measure of the relative strength of the Coulomb interaction, one can use the 
effective fine-structure constant $\alpha_{\text{G}} = e^2/4\pi \epsilon_0 \epsilon \hbar v_F$,
where  $\epsilon=\epsilon_r$ or $(1+\epsilon_r)/2$ is the effective
background dielectric constant for a single layer of graphene embedded in, 
or grown on top of a substrate, respectively. 
As we discuss in this Letter, the Coulomb interaction introduces a sensitivity to the 
dielectric environment via background screening, contradicting the widely believed paradigm that the 
physical properties of graphene are largely independent of the environment.
For freestanding graphene in vacuum $\alpha_{\text{G}} \approx 2.41$, 
indicating prominent Coulomb interaction effects.

It is well known from semiconductor physics that the Coulomb attraction between electrons and 
holes may lead to the formation of bound electron-hole pairs, i.e. excitons.
While excitonic effects in metals are generally believed to be of minor 
importance because of strong screening, in graphene the screening of the long-ranging part of the 
Coulomb interaction is suppressed by the vanishing density of states at the Dirac points.
Hence, it is not surprising that excitonic resonances have been observed in one-dimensional 
metallic carbon nanotubes \cite{Wang2007}.
Excitonic binding energies in these systems ranging from 50 to 100 meV have been calculated by solving 
the Bethe-Salpeter equation and within a density matrix approach \cite{Spataru2004, Malic2010}.
For planar 2D graphene, first-principles calculations have become available only recently \cite{Yang2009}. 
They predict a self-energy correction of the Fermi velocity that is in good agreement with experimental 
findings \cite{Zhang2005} and an excitonic shift of the dominant optical absorption peak as large as 600 meV. 
Hartree-Fock based configuration-interaction calculations for graphene quantum dots 
yield an excitonic redshift of approximately 300 meV \cite{Guclu2010}. 

In this Letter, we show that the relative importance of the Coulomb interaction effects in graphene 
is dominated by its dielectric environment. In particular, we predict that freestanding graphene 
in vacuum has Coulombic properties that are very different from those of epitaxially grown graphene on 
substrates like $\text{SiO}_2$ or SiC with relative $\epsilon$ in the THz range of 
4.45 and 9, respectively. 

A useful and convenient criterion for the importance of Coulomb interaction effects is the existence of bound electron-hole pairs, i.e. excitions. The prediction of a finite exciton binding energy in a gapless system like graphene, is an indication for the fact that  
the system energy an be reduced below the tight binding (TB) ground-state level by exciton formation. In order to investigate this feature, we consider the 
infinitesimal transformation  $U(\beta)={\rm \exp (\beta B_\lambda^\dagger})$ generated by the
 exciton operator
$B^\dagger_\lambda=\sum_{\bk}\phi_\lambda(\bk) e^\dagger_\bk h^\dagger_{-\bk}$. 
where $\phi_\lambda(\bk)$ is a normalized wave function,
and  $e^\dagger_\bk$ and $h^\dagger_{-\bk}$ are electron and hole creation operators defined within the bands.  
Starting from the TB ground state, an infinitesimal transformation yields an energy shift
\begin{eqnarray}
\delta E&=&\beta^2\left(\sum_\bk \varepsilon_{\bk} \,|\phi_{\lambda}(\bk)|^2-\sum_{\bk\bp}\phi_{\lambda}^*(\bk)V^{+}_{\bk,\bp}\, \phi_{\lambda}(\bp)\right.
\nonumber\\
&-&\left.\frac{1}{2}\sum_{\bk\bp}\Bigl[V^{-}_{\bk,\bp}\,\phi_{\lambda}^*(\bk) \phi^*_{\lambda}(\bp)+
V^{-}_{\bk,\bp}\,\phi_{\lambda}(\bk) \phi_{\lambda}(\bp) \Bigr]\right).
\end{eqnarray}
Here, $\varepsilon_{\bk}=\varepsilon^ e_{\bk}+\varepsilon^h_{\bk}$ are the electron and hole dispersion relations
within the TB approximation \cite{Wallace47, Semenoff84, Castro-Neto2009}
and $V^\pm_{\bk,\bp}$ are the Coulomb matrix elements calculated with the electron and hole wave functions\cite{Hirtschulz2008, Gronqvist2010}.  

The shift $\delta E$ is negative if the graphene Wannier equation
\begin{equation} 
\varepsilon_{\bk} \,\phi_{\lambda}(\bk)
-\sum_\bp
\Bigl[V^{+}_{\bk,\bp}\, \phi_{\lambda}(\bp)
+ V^{-}_{\bk,\bp}\, \phi^\star_{\lambda}(\bp) \Bigr]
= E_{\lambda} \, \phi_{\lambda}(\bk) \, ,
\label{Wannierequation}
\end{equation}
has bound state solutions with $E_\lambda<0$. 
Eq. \ref{Wannierequation} is a direct generalization of the well known Wannier-equation in semiconductor physics, 
and describes the relative motion of a Coulomb-interacting electron-hole pair with zero center of mass momentum (c.o.m)\cite{HaugKoch}. 
Altough a general dispersion  
does not allow for the decoupling of the relative and center-of-mass (c.o.m.) motion problems 
\cite{Li2010,Sabio2010}, the smallest pair energy is obtained 
for states with vanishing c.o.m. momentum, as the kinetic c.o.m. energy is always positive. Hence, bound state solutions of Eq. \ref{Wannierequation}
clearly limit the validity range of a weakly interacting model system.
\begin{figure}
\centering{\includegraphics[width=8cm]{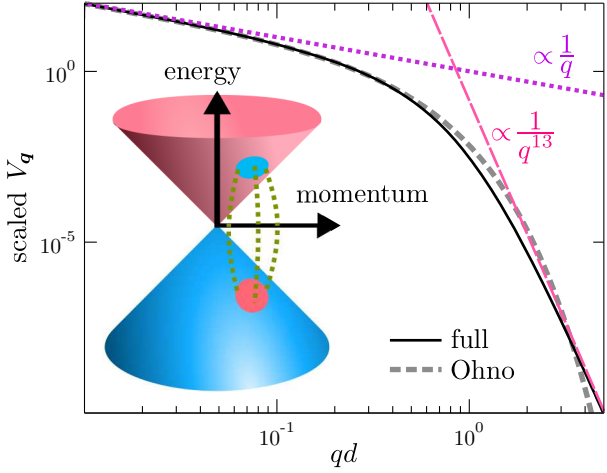}}
\caption{Coulomb matrix element for one sheet of graphene. 
The full Coulomb potential (solid line) and the Ohno potential (short-dashed gray line) 
are shown together with two asymptotic forms (long-dashed and dotted lines). 
Inset: Band structure and a single electron-hole-pair excitation in graphene.}
\label{Coulpot}
\end{figure}
Within the linear approximation, the graphene Wannier equation is equivalent to the Dirac 
two-body problem with zero c.o.m.
Recent real space studies considering the Dirac Coulomb problem 
predict instabilities if the effective fine structure constant exceeds a critical value \cite{Pereira2007, Shytov2007, Wang2010,Sabio2010}.
 Although restricted to positiv energies, the adressed topics
are closely related to the exciton problem and indicate 
a nontrivial dependence on the background dielectric constant.
To characterize the influence of the dielectric environment, 
 we calculate the binding energy of  
electron-hole pairs with zero c.o.m. as a function of the effective background $\epsilon$ by numerically solving the Wannier equation (\ref{Wannierequation}).

In the vicinity of the Dirac points \cite{Wallace47}, we can write the electron-hole 
dispersion as $\varepsilon_{\bk}\approx2\hbar v_F k$. 
The attractive Coulomb interaction is described by the potential 
$V^{\pm}_{\bk,\bp} = V_{\bk-\bp} (1 \pm \cos \theta_{\bk,\bp}) /2 $,
calculated with the tight-binding electron and hole wave functions in the vicinity of the Dirac points, 
similarly to the approach in Refs. \cite{Hirtschulz2008, Gronqvist2010}.
Due to the finite extension of the $p$ orbitals perpendicular to the plane, real graphene is not 
strictly two-dimensional and the 2D Coulomb matrix element $V^{\text 2D}_\bq$ has to be modified 
accordingly. In addition, the dielectric environment of graphene screens the Coulomb interaction 
by the effective background $\epsilon$. As a combined effect, one finds
\begin{equation}
V_\bq = \frac{1}{\epsilon}\, V^{\text{2D}}_\bq  F( qd)
\label{Coulgraphene}\,,
\end{equation}
where the form factor $F(qd)$ depends on the effective graphene thickness $d$. 
The major effect of $F$ is to supply an intrinsic length,
needed to fix the length and energy scale.
As shown in Fig.~\ref{Coulpot}, the Coulomb potential can be approximated quite 
well by an Ohno potential with $F(x)=e^{-5x}$.
For the numerical evaluations, we take $d=0.18$\,\AA\  which is obtained from a 
density functional computation \cite{John}. 
Assuming an $s$-like ground state, we numerically solve Eq.~\eqref{Wannierequation} using the 
linear dispersion and the full graphene Coulomb matrix elements. More specifically, we have 
chosen $v_F = 9.07\cdot 10^5$ m/s, yielding $\alpha_{\text{G}} \approx 2.41$ for graphene in vacuum 
($\epsilon=1$) and vary the effective background dielectric constant $\epsilon$ to change 
$\alpha_{\text{G}} \propto 1/\epsilon$.

\begin{figure}
\centering{\includegraphics[width=8cm]{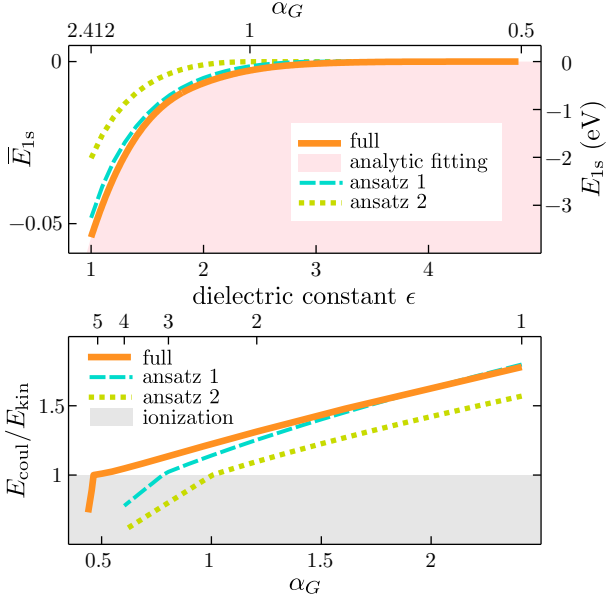}}
\caption{(top) Exciton binding energy vs.~background dielectric constant (lower axis) 
and $\alpha_{\text{G}}$ (upper axis). 
The left axis presents the scaled energies and the right axis gives the energy in eV calculated 
using $d=0.18$\,\AA. The full numerical solution (solid line) is compared with results obtained 
using the ansatz wave functions $f(y)\propto (1+y^2)^{-3/2}$ (dashed line) and 
$f(y) \propto e^{\smash{-y^2}}$ (dotted line). 
The shaded area corresponds to the analytic result \eqref{eq:E_1s-estimate}. 
(bottom) Ratio of the Coulomb and kinetic energy vs.~$\alpha_{\text{G}}$ (lower axis) and $\epsilon$ 
(upper axis). The shaded area indicates the region of ionized solutions.}
\label{evseps}
\end{figure}

We find the very large binding energy $E_{\text{1s}}=-3.8$\,eV for the lowest exciton state of graphene 
in vacuum ($\epsilon=1$, $\alpha_{\text{G}}=2.41$). This energy rapidly decreases with increasing 
$\epsilon$ (decreased $\alpha_{\text{G}}$), e.g. $\epsilon =2$ ($\alpha_{\text{G}}=1.21$), 
3 ($\alpha_{\text{G}}=0.80$), and 4 ($\alpha_{\text{G}}=0.60$) produce 
$E_{\text{1s}}=-0.47$\,eV, $-44$\,meV, and $-0.94$\,meV, respectively. 
As discussed below, these numbers scale strictly as $1/d$ if the effective thickness of 
graphene is changed. Fig.\ \ref{evseps}a shows the full $\epsilon$ (lower scale) and $\alpha_{\text{G}}$ 
(upper scale) dependency of $E_{\text{1s}}$. 
Within the numerical accuracy, we find a critical value $\epsilon_{\text {ion}} \approx 4.8$ 
($\alpha_{\text{ion}} \approx 0.5$).
For $\epsilon< \epsilon_{\text {ion}}$ ($\alpha_{\text{G}} >  \alpha_{\text{ion}}$), 
we always find a negative value for $E_{\text{1s}}$,
indicating the existence of a bound exciton state. Above (below) the threshold value 
our computations produce strictly positive-valued $E_{\text{1s}}$, 
indicating that only ionized electron--hole states exist. 
Hence, our results show that the presence of bound electron--hole pairs, i.e. strong Coulomb 
interaction effects, depends extremely sensitively on the dielectric environment of the graphene.
 
The transition from the strongly to a weakly Coulomb interacting regime can be detected even more 
clearly by studying the ratio between the Coulomb and the kinetic energy computed using the
numerically determined wave functions $\phi_{\text{1s}}(\bk)$. In Fig. \ref{evseps}b, we plot 
$g = E_{\text{coul}}/E_{\text{kin}}$ as function of $\alpha_{\text{G}}$ (lower scale) and $\epsilon$ 
(upper scale). For the bound states, the Coulomb energy dominates the kinetic energy so
that $g$ is larger than one, while the kinetic energy exceeds the Coulomb energy for ionized 
electron--hole pairs. The shaded area in Fig. \ref{evseps}b indicates the region of ionized 
states. We see that $g$ crosses from the region of bound states to the region of ionized states 
roughly at $\epsilon_{\text{ion}}=4.8$ ($\alpha_{\text{ion}}=0.5$), in full agreement with the 
ionization threshold observed in Fig. \ref{evseps}a. 
If we regard $g$ as an order parameter, the exciton binding experiences a second-order phase 
transition at $\alpha_{\text{ion}}$ because  $\partial g/\partial \alpha_{\text{G}}$ becomes 
discontinuous at $\alpha_{\text{G}}=\alpha_{\text{ion}}$.

In order to analyze the graphene Wannier equation further, we introduce the dimensionless quantities
$\bar {\bk}=\bk d$, $\bar E_\lambda=E_\lambda d/2\hbar v_F$, $\bar \phi(\bar k)=\phi(k)/d$,  
and take the continuum limit
$\sum_{\bk}\rightarrow{\cal A}/(2\pi)^2\int \ud^2k$. This yields the dimensionless equivalent to 
Eq.~\eqref{Wannierequation} 
\begin{equation}
\label{Wannierdimensionless}
\bar{k} \bar{\phi}_{\lambda}(\bar{\bk})-
\frac{\alpha_{\text{G}}}{4\pi}
\int \ud^2 {\bar { k'}}\frac{F(|\bar{\bk}-\bar{\bk}'|)}{|\bar {\bk}-\bar{\bk}'|}
\bar {\phi}_{\lambda}(\bar{\bk}')=\bar{E}_{\lambda} \bar{\phi}_{\lambda}(\bar{\bk})
\end{equation}
which has to be solved subject to the normalization 
condition $\int \ud^2\bar {k} |\bar {\phi}_{ \lambda}(\bar { \bk})|^2=1$. 
Equation~\eqref{Wannierdimensionless} shows that the effective graphene fine-structure 
constant $\alpha_{\text{G}} = e^2/4\pi \epsilon_0 \epsilon \hbar v_F$ is the single 
parameter combination characterizing the solution. 
In particular, any solution is independent of the sheet thickness $d$
and hence, valid for all graphene-like systems. The effective thickness is
needed only to fix the length and energy scale. 
Moreover, one recognizes that any solution of Eq.~\eqref{Wannierdimensionless} with a finite eigenvalue
$\bar {E}_\lambda$ produces a divergent physical energy eigenvalue in the strict 2D limit $d\rightarrow 0$.

Assuming $d\neq 0$, and using the dimensionless scaling parameter $u=a/d$, we treat
the energy eigenvalue $\bar E_\lambda$ as functional of the scaling parameter:
\begin{eqnarray} 
\bar{E}_{\lambda} =
\frac{1}{u} \left[T^{\left[1\right]}_\lambda-\alpha_{\text{G}} \,{\textstyle V_\lambda (u)} \right],
\label{Egvsa}
\end{eqnarray}
with 
$ V_\lambda (u) =  \frac{1}{4\pi} \!\int\! \ud^2 y\, \ud^2 y'\,
\psi^*_\lambda(\by;u)\frac{F(|\by-\by'|/u)}{|\by-\by'|}\psi_\lambda(\by';u)$
and $T_\lambda^{\smash{[n]}} = \!\int\! \ud^2 y \,y^n|\psi_\lambda(\by;u)|^2$.
Stationary points of Eq.~\eqref{Egvsa} correspond to either bound states or resonances of the system, 
depending on the sign of $\bar E_\lambda$.
                                        
If it exists, the lowest bound state defines the characteristic length and energy 
scales of the exciton, i.e. the exciton Bohr radius and binding energy, in terms of the sheet thickness $d$ and 
the energy unit $2\hbar v_F/d$. 
Since the wave function scales with the sheet thickness which is independent of $\alpha_{\text{G}}$, 
and the stationary points of Eq.~\eqref{Egvsa} depend on $\alpha_{\text{G}}$ via the wave equation,
the scaled wave function $\psi(\by;u)$ has an implicit nontrivial $u$ dependence. 
As a consequence, $T_\lambda^{\smash{[n]}}$ and the Coulomb integral also depend on $u$.
In contrast, the implicit $u$ dependence of the total energy
vanishes for any solution of the wave equation.
Hence, stationary points of Eq.~\eqref{Egvsa} are found for 
$\ud\bar E_\lambda/\ud u = -\bar E_\lambda/u - \alpha_{\text{G}} u^{-1}\, \partial V(u)/\partial u=0$. 
Since $V(u)$ is a monotonically increasing function of $u$, this predicts stable bound states
for any value of $\alpha$ with $\bar{E}_{\lambda} < 0$.
From these arguments, we obtain the condition 
$\alpha_{\text{G}}>\alpha_{\text{ion}} = T^{\smash{[1]}}_\lambda / V_\lambda (u)$.
For every regular, normalized wave function, $T^{\smash{[1]}}>0$ and $V(u)$ does not diverge, 
and hence, graphene has a finite ionization threshold $\alpha_{\text{ion}}$. This value 
separates the strong Coulomb-interaction regime where bound states exist from the weakly 
interacting configurations without bound states.

An analytical estimate for $\alpha_{\text{ion}}$ can be obtained using a variational 
analysis with a trial wave function $\psi(\by;u)\rightarrow f(y)$ with fixed width to calculate 
the ratio $g = E_{\text{coul}}/E_{\text{kin}}$ of the Coulombic and kinetic energy contributions. 
Figure \ref{evseps} shows the results for $f(y)=\sqrt{8 \pi}(1+y^2)^{-3/2}$ with the ionization 
threshold $\alpha_{\text{ion}}=\pi/4$ (dashed line) and $f(y)=\sqrt{8 \pi} e^{\smash{-y^2}}$ with
$\alpha_{\text{ion}}=1$ (dotted line) as function of $\alpha_{\text{G}}$. 
We notice that the resulting ionization values of the fine-structure constant depend on the 
explicit form of the trial wave function. Clearly, these estimates only provide an upper bound 
since the true wave function may vary its shape via the $u$ dependence to minimize the energy.

The same analysis can be repeated for a model system with a parabolic electron--hole dispersion 
$\varepsilon_{\bk} = \hbar^2 {\bk}^2/2 m$ parametrized by the mass $m$. This situation 
corresponds to a two-band model for an effectively 2D quantum well system. Here, we obtain
$ E^{\text{sc}}_{\text{1s}} = \frac{\hbar^2}{2m a^2} \bigl[ T^{\left[2\right]}
- \frac{a}{a_0}  V(d/a) \bigr]$
where $a_0=2 \pi \hbar^2\epsilon\epsilon_0/me^2$ assigns a new length scale. Whereas this cannot be 
done for graphene, for the parabolic model we may now set $d=0$ to proceed to the ideal 2D case. 
In this situation, the kinetic energy scales like $1/a^2$ while the Coulomb energy is proportional 
to $1/a$.
The variational condition $\partial E^{\text{sc}}_{\text{1s}}/\partial a=
-E/a-E_{\text{kin}}/a=-2E/a + E_{\text{coul}}/a=0$ yields the well known 
virial theorem, the Bohr radius $a=a_0$, and the binding energy 
$E^{\text{sc}}_{\text{1s}} = - \frac{\hbar^2}{2 m a_0^2}\,T^{[2]}$ for any bound state solution 
regardless of $\epsilon$. Hence, there is no ionization threshold in parabolic-band semiconductors.

Returning to graphene, we now derive an expression which explicitly shows the extreme 
sensitivity of $E_{\text{1s}}$ on $\epsilon$ 
(or $\alpha_{\text{G}}$). For this purpose, we combine the properties resulting from Eq.~\eqref{Egvsa} 
with the numerical results. A direct differentiation of Eq.~\eqref{Egvsa} produces 
$\partial \bar{E}_{\text{1s}}/\partial \alpha_{\text{G}} 
= \frac{\partial \bar{E}_{\text{1s}}}{\partial u} 
\frac{\partial u}{\partial \alpha_{\text{G}}} -  V(u)/u
= - \bar{E}_{\text{coul}}/\alpha_{\text{G}}$ 
since $\partial \bar{E}_{\text{1s}}/\partial u$ 
vanishes for the ground state.
As suggested by Fig.~\ref{evseps}, we can use 
$\bar{E}_{\text {coul}}/\bar{E}_{\text{kin}} - 1 
= c(\alpha_{\text{G}} -\alpha_{\text{ion}})$. 
A straightforward integration gives
\begin{eqnarray} 
\bar{E}_{\text{1s}}
= \frac{\bar{E}^{\text{vac}}_{\text{1s}}}{\epsilon}\,
\left( \frac{\epsilon_{\text{ion}} - \epsilon}{\epsilon_{\text{ion}}-1}
\right)^P\,,\quad \epsilon \le \epsilon_{\text{ion}}
\label{eq:E_1s-estimate}\,,
\end{eqnarray}
where $\bar{E}^{\text{vac}}_{1s}$ is the binding energy of graphene in vacuum and 
$P=1/ c\alpha_{\text{ion}}$.

Based on Fig.~\ref{evseps}, we use $P=4.5$, $\epsilon_{\text{ion}}=4.8262$, and 
$\bar{E}^{\text{vac}}_{1s}=-0.0542$. In particular, we see that $\bar{E}_{1s}$ scales like 
$(\epsilon_{\text{ion}}-\epsilon)^{4.5}/\epsilon \rightarrow (\epsilon_{\text{ion}}-\epsilon)^{4.5}$ 
close to the ionization threshold which explains the rapid decrease of the binding energy 
before the ionization is reached.
The shaded area in Fig.~\ref{evseps} shows the analytic estimate of $\bar{E}_{1s}$ 
together with the numerically obtained binding energy (solid line), nicely demonstrating that 
Eq.~\eqref{eq:E_1s-estimate} captures the essential features.

In conclusion, our calculations predict an extreme sensitivity of the Coulomb effects in graphene 
on the dielectric properties of the environment. We find strongly bound excitons for freestanding 
graphene in vacuum, whereas all bound states disappear for the effective background dielectric 
constant $\epsilon_{\rm ion}=4.8$ which corresponds to the graphene fine-structure constant 
$\alpha_{\text{G}}=1/2$. Our results are in general agreement with an analytical
real space analysis of the 2D Dirac two-body problem \cite{Sabio2010}, 
studies of the Coulomb scattering at an impurity charge $Ze$ in 
graphene \cite{Pereira2007, Shytov2007, Wang2010}, 
as well as Monte Carlo calculations \cite{Drut2009,Drut2009b}, predicting instabilities for critical 
$\alpha$ values in the range of $0.5$ to $1.66$.
Physically, this situation is realized, e.g., when graphene is deposited onto substrates or sandwiched 
between dielectric media such that the effective background dielectric constants exceeds 4.8. 
While in the regime of weak Coulomb interaction quasi-free electrons and holes are responsible 
for the low-energy electronic and optical properties, those are dominated by bound 
excitons for small background dielectric constants. 
On this basis, we expect a pronounced influence of the dielectric environment on the graphene 
ground state and on properties like the low-density conductivity, the quantum Hall effect, 
or the terahertz response.

Acknowledgements: The authors thank Prof.\ M.\ Kira, Marburg, for many stimulating discussions and Prof.\ J.\ Sipe for the information 
on his DFT results prior to publication.


\begin{thebibliography}{Yang2009}

\bibitem{Wallace47} P.R. Wallace, Phys. Rev. {\bf 71}, 622, (1947).

\bibitem{Semenoff84} G.W. Semenoff, Phys. Rev. Lett. {\bf 53}, 2449 (1984).

\bibitem{Castro-Neto2009} A.H. Castro Neto, F. Guinea, N.M.R. Peres, K.S. Novoselov, and A.K. Geim, 
Rev. Mod. Phys. {\bf 81}, 109 (2009).

\bibitem{Novoselov2005} K.S. Novoselov, A.K. Geim, S.V. Morozov, D. Jiang, J.I. Katsnelson, 
I.V. Grogorieva, S.V. Dubonos, and A.A. Firsov, Nature {\bf 438}, 197 (2005).

\bibitem{Zhang2005} Y. Zhang, Y.-W. Tan, H.L. Stormer, and P. Kim, Nature {\bf 438}, 201 (2005).

\bibitem{Deacon2007} R.S. Deacon, K.-C. Chuang, R.J. Nicholas, K.S. Novoselov, and A.K. Geim, 
Phys. Rev. B {\bf 76} R081406 (2007).

\bibitem{Jiang2007} Z. Jiang, E.A. Henriksen, L.C. Tung, Y.-J. Wang, M.E. Schwartz, M.Y. Han, 
P. Kim, and H.L. Stormer, Phys. Rev. Lett. {\bf 98}, 197403 (2007).

\bibitem{Zhou2006} S.Y. Zhou, G.-H. Gweon, J. Graf, A.V. Federov, C.D. Sparatu, R.D. Diehl, 
Y. Kopolevich, D.-H. Lee, S.G. Louie, and A. Lanzara, Nat. Phys. {\bf 2}, 595 (2006).

\bibitem{Bostwick2007} A. Bostwick, T. Ohta, T. Seyller, K. Horn, and E. Rotenberg, 
Nat. Phys {\bf 3}, 36 (2007).

\bibitem{Gonzalez99} J. Gonz\' alez, F. Guinea, M.A.H. Vozmediano, Phys. Rev. B {\bf 59}, R2474 (1999).


\bibitem{Sheehy2007} D.E. Sheehy and J. Schmalian, Phys. Rev. Lett. {\bf 99}, 226803 (2007).

\bibitem{Fritz2008} L. Fritz, J. Schmalian, M. M\" uller, and S. Sachdev, Phys. Rev. B {\bf 78}, 
085416 (2008).

\bibitem{Sinner2010} A. Sinner and K. Ziegler, Phys. Rev. B {\bf 82}, 165453 (2010).

\bibitem{Khveshchenko2001} D.V. Khveshchenko, Phys. Rev. Lett. {\bf 87}, 246802 (2001).

\bibitem{Khveshchenko2006} D.V. Khveshchenko and W.F. Shively, Phys. Rev. B {\bf 73}, 115104 (2006).

\bibitem{Juricic2009} V. Juricic, I.F. Herbut, and G.W. Semenoff, Phys. Rev. B {\bf 80}, R081405 (2009).

\bibitem{Drut2009} J.E. Drut and T.A. L\"ahde, Phys. Rev. Lett. {\bf 102}, 026802 (2009).

\bibitem{Drut2009b} J.E. Drut and T.A. L\"ahde, Phys. Rev. B {\bf 79}, 165425 (2009).

\bibitem{Guclu2010} A.D. G\"ucl\"u, P. Potasz, and P. Hawrylak, Phys. Rev. B {\bf 82}, 155445 (2010).

\bibitem{Yang2009} L. Yang, J. Deslippe, C.-H. Park, M.L. Cohen, and S.G. Louie, 
Phys. Rev. Lett. {\bf 103}, 186802 (2009).

\bibitem{Reed2010} J.P. Reed, B. Uchoa, Y.I. Joe, Y. Gan, D. Casa, E. Fradkin, and P. Abbamonte,
Science {\bf 330}, 805 (2010).

\bibitem{Malic2010} E. Malic, J. Maultzsch, S. Reich, and A. Knorr, Phys. Rev. B {\bf 82}, 
035433 (2010).

\bibitem{Min2008} H. Min, R. Bistritzer, J.-J. Su, and A. H. MacDonald, Phys. Rev. B {\bf 78}, 121401(R) (2008).

\bibitem{Berman2008a} O. L. Berman, Y. E. Lozovik, and G. Gumbs, Phys. Rev. B {\bf 77}, 155433 (2008).

\bibitem{Berman2008b} O. L. Berman, R. Ya. Kezerashvili, and Y. E. Lozovik, Phys. Rev. B {\bf 78}, 035135 (2008).


\bibitem{Wang2007} F. Wang, D.J. Cho, B. Kessler, J. Deslippe, P.J. Schuck, S.G. Louie, 
A. Zettl, T.F. Heinz, and R. Shen, Phys. Rev. Lett. {\bf 99}, 227401 (2007).

\bibitem{Spataru2004} C.D. Spataru, S. Ismail-Beigi, L.X. Benedict, and S.G. Louie, 
Phys. Rev. Lett. {\bf 92}, 077402 (2004).

\bibitem{Hirtschulz2008} M. Hirtschulz, F. Milde, E. Malic, S. Butscher, C. Thomsen,  
S. Reich, and A. Knorr, Phys. Rev. B {\bf 77}, 035403 (2008)

\bibitem{Gronqvist2010} J. H. Gr{\"o}nqvist, M. Hirtschulz, A. Knorr, and M. Lindberg, 
Phys. Rev. B {\bf81}, 035414 (2010)

\bibitem{HaugKoch} H. Haug and S.W. Koch, {\it Quantum Theory of the Optical and Electronic 
Properties of Semiconductors}, World Scientific Publishing Co. Pte. Ltd., Singapore, 5th Ed. (2009).
 
\bibitem{Li2010} X. Li, E. A. Barry, J. M. Zavada, M. Buongiorno Nardelli, and K. W. Kim, 
Appl. Phys. Lett. {\bf 97}, 082101 (2010)
\bibitem{Sabio2010} J. Sabio, F. Sols, and F. Guinea, Phys. Rev. B {\bf 81}, 045428 (2010)
  


\bibitem{Shytov2007} A.V. Shytov, M.I. Katsnelson, and L.S. Levitov, Phys. Rev. Lett. {\bf 99}, 236801 (2007)

\bibitem{Pereira2007} V.M. Pereira, J. Nilsson, and A.H. Castro Neto, Phys. Rev. Lett. {\bf 99}, 166802 (2007).

\bibitem{Wang2010} J. Wang, H. A. Fertig, and G. Murthy, Phys. Rev. Lett. {\bf 104}, 186401 (2010)




\bibitem{John} 
This is taken from an all-electron ground state LDA calculation (Perdew-Wang functional) done at 
the PAW level, using the ABINIT program; a supercell calculation was used to study an isolated 
graphene sheet in vacuum. The thickness of the graphene sheet, $d$, was deduced from the width of the found 
ground state charge density. We thank Cuauhtemoc Salazar and J.E. Sipe for providing us with 
these results.     
         
\bibitem{Rana2007} F. Rana, Phys. Rev. B{\bf 76}, 155431 (2007).
\end{thebibliography}
\end{document}